\documentclass[jgr]{agu2001}
\lefthead{BRUNO ET AL.}
\righthead{SOLAR WIND INTERMITTENCY}
\authoraddr{R. Bruno and B. Bavassano, Istituto di Fisica dello Spazio
Interplanetario, CNR, Via del Fosso del Cavaliere 100, 00133 Rome, Italy. (e-mail: bruno@ifsi.rm.cnr.it ;
bavassano@ifsi.rm.cnr.it)}
\authoraddr{V. Carbone, L. Sorriso--Valvo, Dipartimento di Fisica Universit\`a della Calabria, 87036
Rende (Cs), Italy}
\begin{document}

\title{Radial evolution of solar wind intermittency in the inner heliosphere}

\author{R. Bruno}
\affil{Istituto di Fisica dello Spazio Interplanetario, Consiglio Nazionale delle Ricerche, Rome, Italy}
\author{V. Carbone, L. Sorriso--Valvo}
\affil{Dipartimento di Fisica Universit\`a della Calabria, 87036 Rende (Cs), Italy}
\author{B. Bavassano}
\affil{Istituto di Fisica dello Spazio Interplanetario, Consiglio Nazionale delle Ricerche, Rome, Italy}

\input epsf



\begin{abstract}
We analyzed intermittency in the solar wind, as observed on the ecliptic plane, looking at magnetic field and
velocity fluctuations between 0.3 and 1 AU, for both fast and slow wind and for compressive and directional
fluctuations. Our analysis focused on the property that probability distribution functions of a fluctuating field
affected by intermittency become more and more peaked at smaller and smaller scales. Since the peakedness of a
distribution is measured by its flatness factor we studied the behavior of this parameter for different scales to
estimate the degree of intermittency of our time series. We confirmed that both magnetic field and velocity
fluctuations are rather intermittent and that compressive magnetic fluctuations are generally more intermittent
than the corresponding velocity fluctuations. In addition, we observed that compressive fluctuations are always
more intermittent than directional fluctuations and that while slow wind intermittency does not depend on the
radial distance from the sun,  fast wind intermittency of both magnetic field and velocity fluctuations clearly
increases with the heliocentric distance.

We propose that the observed radial dependence can be understood if we imagine interplanetary fluctuations made of
two main components: one represented by coherent, non propagating structures convected by the wind and, the other
one made of propagating, stochastic fluctuations, namely Alfv\'{e}n waves. While the first component tends to
increase the intermittency level because of its coherent nature, the second one tends to decrease it because of its
stochastic nature. As the wind expands, the Alfv\'{e}nic contribution is depleted because of turbulent evolution
and, consequently, the underlying coherent structures convected by the wind, strengthen further on by
stream--stream dynamical interaction, assume a more important role increasing intermittency, as observed.
Obviously, slow wind doesn't show a similar behavior because Alfv\'{e}nic fluctuations have a less dominant role
than within fast wind and the Alfv\'enicity of the wind has already been frozen by the time we observe it at 0.3
AU. Finally, our analysis suggests that the most intermittent magnetic fluctuations are distributed along the
local interplanetary magnetic field spiral direction while, those relative to wind velocity seem to be located
along the radial direction.
\end{abstract}

\begin{article}

\section{1. Introduction}

The basic view that we have of the solar wind is that of a magnetofluid pervaded by fluctuations over a wide range
of scales which are strongly modified by the effects of the dynamics during the expansion into the interplanetary
medium. These effects are more relevant within the inner heliosphere and on the Ecliptic where the stream--stream
dynamics more strongly reprocesses the original plasma and the large velocity shears add new fluctuations to the
original spectrum (Coleman, 1968, Roberts et al., 1991). This scenario has reconciled the "wave" point of view
proposed by Belcher and Davis (1971), i.e. solar origin of the fluctuations, and the "turbulence" point of view,
i.e. local generation due to velocity shears, proposed by Coleman(1968). The first consequence of this scenario is
that large fluctuations of solar origin containing energy interact non--linearly with other fluctuations of local
origin giving rise to an energy exchange between different scales, which can be interpreted as the usual energy
cascade towards smaller scales in fully developed turbulence. As a matter of fact, spacecraft observations have
shown that the spectral slope of the power spectrum of these fluctuations changes with the radial distance from
the sun (Bavassano et., 1982, Denskat and Neubauer, 1983). This behaviour was recognized (Tu et al., 1984) as a
clear experimental evidence that cascade processes due to non--linear interaction between opposite propagating
Alfv\'en waves were active in the solar wind with. One of the consequences of this radial evolution was the
observed radial decrease of the correlation of velocity and magnetic field fluctuations (generally known as cross
helicity, or Alfv\'enicity)(Roberts et al. 1987). These observations finally answered to the question of whether
the observed fluctuations were remnants of coronal processes or were dynamically created during the expansion.
However, successive theoretical models (See review by Tu and Marsch, 1995) which tried to obtain the radial
spectral evolution of the solar wind fluctuations had to deal with peculiarities of the observations that they
could not reproduce within the framework of solely non--linear interacting waves. The lack of a strict
self--similarity of the fluctuations and the consequent non applicability of strict scale invariance (Marsch and
Liu, 1993), the strong anisotropy shown by velocity and field fluctuations (Bavassano et al., 1982, Tu et al.,
1989, Roberts, 1992), the different radial evolution of the minimum variance direction for magnetic field and
velocity (Klein et al., 1993), the lack of equipartition between magnetic and velocity fluctuations (Matthaeus and
Goldstein, 1982, Bruno et al., 1985) all contributed to suggest the idea that fluctuations could possibly be due
to a mixture of propagating waves and static structures convected by the wind. Some kind of filamentary structure,
similar to flux tubes, was firstly proposed by McCracken and Ness (1966) and the observed spectral radial
evolution of the large scale fluctuations has been attributed to the interaction of outward propagating Alfv\'en
waves with these structures (Tu and Marsch, 1993, Bruno and Bavassano, 1991; Bavassano and Bruno, 1992).
Incompressible magnetic structures were found by Tu and Marsch (1991)and magnetic fluctuations with a large
correlation length parallel to the ambient magnetic field, suggested the idea of a quasi--two--dimensional,
incompressible turbulence for which $\vec{k}\cdot\vec{B}=0$
(Matthaeus et al., 1990). Thus, solar wind fluctuations
are not isotropic and scale--invariant, two of the fundamental hypotheses at the basis of K41 Kolmogorov's theory
(1941). This theory is based on an important statistical relation, which characterizes turbulent flows, between
velocity increments $\delta v_r=<|\vec{V}(\vec{x}+\vec{r})-\vec{V}(\vec{x})|>$, measured along the flow direction
$x$, and the energy transfer rate $\epsilon$ at the scale separation $r=|\vec{r}|$, that is $ \delta
v_r\sim(\epsilon r)^{1/3}$ or, more in general, $\delta v^p_r\sim(\epsilon r)^{p/3}$. If $\epsilon$ is constant,
the previous relation simply reads $\delta v^p_r\sim r^{p/3}$ and fluctuations are said to be
\textit{self--similar}, and our signal is a simple fractal. However, as remarked by Landau (Kolmogorov, 1962,
Obukhov, 1962), if $\epsilon$ statistically depends on scale due to the mechanism that transfers energy from
larger to smaller eddies, $\epsilon$ will be replaced by $\epsilon_r$ and a new scaling has to be evaluated
$\delta {v^p_r}\sim r^{p/3}<\epsilon_r^{p/3}>$. Expressing $\epsilon_r^{p/3}$ via a scaling relation with $r$, we
obtain $<\epsilon_r^{p/3}>\sim r^{\tau_{p/3}}$ and, consequently, $\delta{v^p_r}\sim r^{s_p}$ where
$s_p=p/3+\tau_{p/3}$ is generally a nonlinear function of $p$. This means that the global scale invariance
required in the K41 theory would release towards a local scale invariance where different fractal sets
characterized by different scaling exponents can be found.

One of the consequences of this lack of a universal scale invariance, directly observed in experimental tests, is
that the shape of the probability density functions (PDFs) of the velocity increments at a given scale is not the
same for each scale but roughly evolves from a Gaussian shape, near the integral scale, to a distribution whose
tails are much flatter than those of a Gaussian, resembling a stretched exponential near the dissipation scale.
This means that the largest events, contained in the tails of the distribution, do not follow the Gaussian
statistics but show a much larger probability. This phenomenon is also called intermittency and, in practice,
fluctuations of a generic time series affected by intermittency, alternate intervals of very high activity to
intervals of quiescence.

Because of this lack of Gaussianity, the study of the fluctuations based on conventional spectral analysis is
strongly limited, and the second order moment of the distribution is not longer the limiting order. An alternative
way for characterizing the fluctuations is to investigate directly the differences of a fluctuating field over all
the possible spatial scales and look at moments of orders higher than 2, adopting the so-called multifractal
approach (Parisi and Frisch, 1985). A convenient statistical tool to perform this study is the so--called $p-th$
order structure function (SF) defined as $S^p_r=<|\vec{V}(\vec{x}+\vec{r})-\vec{V}(\vec{x})|^p>$ and $S^p_r$ is
expected to scale as $r^{s_p}$. SFs are then computed for various orders as a function of all the possible scales
and each order provides a value of the scaling exponent $s_p$. If observations show a non--linear departure from
the simple $s_p=p/3$ (or $s_p=p/4$ for the MHD case (Carbone, 1993)) this is an indication that intermittency is
present. This method was introduced for the first time in space plasma studies by Burlaga (1991) who studied the
exponents $s_p$ of structure functions based on Voyager's observations of solar wind speed at 8.5 AU. This author
found that, similarly to what is found in ordinary laboratory turbulent fluids, the exponent $s_p$ was not equal
to $p/3$, as expected in the K41 theory. This exponent was found to scale non-linearly with the order $p$ and to
be consistent with a variety of newer theories of intermittent turbulence, including Kolmogorov--Obukhov (1962).
The first results obtained by Burlaga (1991) and Carbone et al. (1995) not only revealed the intermittent
character of interplanetary magnetic field and velocity fluctuations but also showed an unexpected similarity to
those obtained for laboratory turbulence (Anselmet et al.,1984). These results showed consistency between
observations on scales of 1 AU and laboratory observations on scales of meters, suggesting a sort of universality
of this phenomenon, which was independent on scale.

While previous results referred to observations in the outer heliosphere, Marsch and Liu (1993) firstly
investigated solar wind scaling properties in the inner heliosphere. For the first time they provided some
insights on the different intermittent character of slow and fast wind, on the radial evolution of intermittency
and on the different scaling characterizing the three components of velocity. They also concluded that the
Alfv\'{e}nic turbulence observed in fast streams starts from the Sun as self--similar but then, during the
expansion, decorrelates becoming more multifractal. This evolution was not seen in the slow wind supporting the
idea that turbulence in fast wind is mainly made of Alfv\'en waves and convected structures (Tu and Marsch, 1993)
as already inferred by looking at the radial evolution of the level of cross--helicity in the solar wind (Bruno
and Bavassano, 1991). As we will see in the following, although the tools used in our analysis differ from those
used by Marsch and Liu (1993) our results fully confirm their results but also add some more inferences on the
radial evolution of solar wind intermittency.

Successively, several other papers tried to understand the phenomenon of intermittency in the solar wind looking
for the best model which could fit the observations or could establish whether the observed scaling was closer to
that shown by an ordinary fluid or rather by a magnetofluid as predicted by Kolmogorov (1941) and Kraichnan
(1965), respectively. Ruzmainkin et al., (1995) studying fast wind data observed by Ulysses developed a model of
Alfv\'enic turbulence in which they reduced the spectral index of magnetic field fluctuations by an amount
depending on the intermittency exponent. They found a close agreement with the expected Kraichnan scaling for a
magnetofluid ($3/2$) and concluded that their results were consistent with a turbulence based on random--phased
Alfv\'en waves (Kraichnan, 1965).

Tu et al., (1996) re--elaborated the Tu (1988) model of developing
turbulence including intermittency derived from the p-model of
Meneveau and Sreenivasan (1987). They obtained a new expression
for the scaling exponent that took into account that, for
turbulence not fully developed, the spectral index is not defined
yet.

Carbone et al., (1995), for the first time adopted the Extended
Self--Similarity (ESS) concept (Benzi et al., 1993) to
interplanetary data collected by Voyager and Helios, and looked
for differences in the scaling properties between interplanetary
magnetofluid and ordinary fluid turbulence obtained in
laboratory. ESS is a powerful method to easily recover the
scaling exponent of the fluctuations exploiting the
interdependency of the structure functions of various orders.
These authors concluded that, differences exist between scaling
exponents in ordinary (unmagnetized) fluid flows and
hydromagnetic flows.

Horbury and Balogh (1997) performed a comprehensive structure
function analysis of Ulysses data and concluded that
interplanetary magnetic field fluctuations are more
Kolmogorov--like rather than Kraichnan--like.

Veltri and Mangeney (1999), adopting a method based on the
discrete wavelet decomposition of the signal identified for the
first time intermittent events. Successively, using conditioned
structure-functions, they excluded any contribution from
intermittent samples and were able to recover the scaling
properties of the MHD fluctuations. In particular, the radial
component of the velocity displayed the characteristic Kolmogorov
slope $s_p=p/3$ while the other components displayed the Kraichnan
slope $s_p=p/4$.

All previous works dealt with the scaling exponents $s_p$ of the structure functions $S^p_r$, aiming to show that
they follow an anomalous scaling with respect to that expected from K41 theory for turbulent fluids. This
anomalous scaling is strictly related to the way Probability Distribution Functions (PDFs) of the increments
change with scale. It is interesting to notice that if we consider fluctuations that follow a given scaling, say
$\delta v_r=<|v(x+r)-v(x)|>\sim r^h$ and introduce a change of scale, say $r\rightarrow \ell  r$ ($\ell
>0$), we end up with the following transformation
$\delta v_{\ell  r}\sim \ell ^h \delta v_r$. The importance of this relation is that the statistical properties of
the left and right--hand--side members are the same (Frisch, 1995), i.e. $PDF(\delta v_{\ell  r})=PDF(\ell
^h\delta v_r)$. This means that if $h$ is unique, the PDFs of the standardized variables $\delta u_r(x) =
(v(x+r)-v(x))/<(v(x+r)-v(x))^2>^{1/2}$ reduces to a unique PDF highlighting the self--similar (fractal) nature of
the fluctuations. In other words, if all the PDFs of standardized fluctuations $\delta u_r$ collapse to a unique
PDF, fluctuations are not intermittent. Intermittency implies multifractality and, as a consequence, an entire
range of values for $h$. Castaing et al. (1990) developed a model based on the idea of a log--normal energy
cascade and showed that the non-Gaussian behavior of the Probability Distribution Functions (PDF's) at small
scales can be represented by a convolution of Gaussians whose variances are distributed according to a log-normal
distribution whose width is represented, for each scale $r$, by the parameter $\lambda^2(r)$. This model has been
adopted, for the first time in the solar wind context, by Sorriso et al., (1999) to fit the departure from a
Gaussian distribution of the PDFs of solar wind speed and magnetic field fluctuations at small scales. As a matter
of fact, Marsch and Tu (1994) had already shown that the PDFs closely resemble a Gaussian distribution at large
scales but, at smaller scales, their tails become more and more stretched as result of the fact that large events
have a probability to happen larger than for a Gaussian distribution.
Their results showed that values of $\lambda^2(r)$ relative to magnetic field were higher than those relative to
velocity throughout the inertial range, confirming that PDF's of magnetic field fluctuations are less Gaussian
than those relative to wind speed fluctuations (Marsch and Tu, 1994). The same authors determined also the
codimension of the most intermittent magnetic and velocity structures, suggesting that within slow wind
intermittency is mainly due to compressive phenomena. Moreover, the use of techniques recently adopted in the
context of solar wind turbulence (Veltri and Mangeney, 1999, Bruno et al. 2001) based on wavelet decomposition
allowed to identify those events causing intermittency. Those events were identified as either compressive
phenomena like shocks or planar sheets like tangential discontinuities separating contiguous regions characterized
by different total pressure and bulk velocity, possibly associated to adjacent flux--tubes.

Lately, Padhye et al., (2001) used the Castaing approach to
describe directly the PDFs of the fluctuations of the overall
interplanetary magnetic field components.  These authors
concluded that all the components followed a rather Gaussian
statistics but they were not able to relate their results to
those obtained by Marsch and Tu (1994) and Sorriso et al. (1999)
who compared PDFs for different time scales. As a matter of fact,
Padhye and co-workers referred to fluctuations respective to the
mean field and not increments as it was done in the previous
mentioned studies and in the present study.

In this paper, we base our analysis on the concept of intermittency as given by Frisch (1995), following which a
random function $v(t)$ is said to be intermittent if the flatness
\begin{equation}\label{frisch}
{\mathcal{F}}=\frac{<(\delta v(t))^4>}{<({\delta v(t))^2>^2}}
\end{equation}
\noindent grows without bound as we filter out the lowest frequency components of our signal and consider only
smaller and smaller scales. Thus, we will define a given time series to be intermittent if $\mathcal{F}$
continually grows at smaller scales and, we will define the same time series to be more intermittent if
$\mathcal{F}$ grows faster. Moreover, if $\mathcal{F}$ remains constant within a certain range of scales, it will
indicate that those scales are not intermittent but simply self--similar and, a value of ${\mathcal{F}} \ne 3$ (3
is the value expected for a Gaussian) would simply indicate that those scales do not have a Gaussian statistics.
This is a simpler way than that used by Sorriso et al. (1999) to look at the behaviour of the flatness to infer the
intermittency character of the fluctuations but, what we gain in simplicity we loose in effectiveness to quantify
the degree of intermittency and, we will only be able to evaluate whether a given sample is more or less
intermittent than another one. In the following sections we will analyze and discuss the radial evolution of
intermittency in the inner heliosphere and on the ecliptic plane evaluating the behavior of $\mathcal{F}$ as
previously illustrated.

\section{2. Data Analysis}

The present analysis was performed using plasma and magnetic field data recorded by Helios 2 during its first
solar mission in 1976 when the s/c repeatedly observed the same corotating stream at three different heliocentric
distances on the ecliptic plane, during three consecutive solar rotations. In order to compare intermittency
between high and low speed plasma, low speed regions ahead of each corotating high speed stream, were also
studied. The three streams, named "1", "2" and "3", respectively, can be identified in Figure~1 where the wind
speed profile and the spacecraft heliocentric distance are shown for the whole Helios 2 primary mission to the
Sun. The exact location of the selected intervals, lasting 2 days each, is shown by the rectangles drawn on the
data profile. Beginning and end of each time interval are shown in Table 1 where, we also show the average
heliocentric distance, the average wind speed, the angle $\hat{\alpha}_{B-V}$ between magnetic and velocity
vectors, the angle $\hat{\beta}_{B-R}$ between magnetic field vector and radial direction and the angle
$\hat{\gamma}_{V-R}$ between velocity vector and radial direction. While the velocity vector is always closely
aligned with the radial direction, magnetic field vector generally follows the expected Archimedean spiral
configuration although this agreement is larger during fast wind than during slow wind time intervals. The data
set is made of 81 sec averages of magnetic and plasma observations recorded in Solar--Ecliptic reference system SE
where, the $\mathcal{X}$ axis is oriented towards the sun, the $\mathcal{Y}$ axis lies on the ecliptic and it is
oriented opposite to the s/c direction of motion and, the $\mathcal{Z}$ axis completes the right--handed reference
system. These fast wind streams are notorious for being dominated by Alfv\'en waves and have been widely studied
since they offer a unique opportunity to observe the radial evolution of MHD turbulence within the inner
heliosphere (for a rather complete review of existing literature related to this topic see Tu and Marsch, 1995).

The aim of the present study is to investigate the behavior of magnetic field and wind velocity intermittency as a
function of heliocentric distance and type of wind (i.e. fast and slow). Although intermittency refers to the
statistical behavior of the fluctuations in the spatial domain, it can be estimated from measurements made in the
temporal domain simply adopting the Taylor's frozen--in hypothesis. This assumption, which is fully acceptable
within the usual conditions of strongly supersonic and super--Alfv\'enic solar wind, allows to treat, with good
approximation, each fluctuation as an eddy and spatial $r$ and temporal $\tau $ coordinates can be mutually
exchanged via the relation $r=V_{sw}\cdot\tau$
where $V_{sw}$ is the solar wind bulk speed. In order to study
intermittency we computed the following estimator of the flatness factor $\mathcal{F}$
\begin{equation}\label{flatness}
{\mathcal{F}}(\tau)=\frac{<S^4_{\tau}>}{<S^2_{\tau}>^2}
\end{equation}\noindent
where ${\tau}$ is the scale of interest and $S^p_{\tau}=<|V(t+\tau)-V(t)|^p>$
is the SF of order $p$ of the generic function $V(t)$. This
definition slightly differs from that given by Frisch (1995)
since we compute the factor $\mathcal{F}$ for each single scale
while Frisch calculates $\mathcal{F}$ using a high--pass filter
whose cutoff  frequency is repeatedly shifted towards higher and
higher frequencies each time. However, in both cases a given
function is considered intermittent if the factor $\mathcal{F}$
increases when considering smaller and smaller scales or,
equivalently, higher and higher frequencies.

A vector field, like velocity and magnetic field, encompasses two
distinct contributions, a compressive one due to intensity
fluctuations that can be expressed as

\begin{equation}\label{compressive}
\delta|\vec{B}(t,\tau)|=|\vec{B}(t+\tau)|-|\vec{B}(t)|
\end{equation} and a directional one due to changes in
the vector orientation

\begin{equation}\label{directional}
\delta\vec{B}(t,\tau)=\sqrt{\sum_{i=x,y,z}(B_i(t+\tau)-B_i(t))^2}
\end{equation}

Obviously, relation \ref{directional} takes into account also
compressive contributions and the expression
$\delta\vec{B}(t,\tau)\geq|\delta|\vec{B}(t,\tau)||$ is always
true.

In the following we will study the flatness factor $\mathcal{F}$ obtained from SFs computed for both compressive
$\xi^p_{\tau}=<|\delta|\vec{B}(t,\tau)||^p>_t$ and directional $\zeta^p_{\tau}=<(\delta\vec{B}(t,\tau))^p>_t$
fluctuations. As regards this last quantity, we like to stress that we verified that magnetic sector changes do
not appreciably influence its value and that in this study only interval (72:00--74:00) contains a magnetic sector
change. Comparing the radial dependence of these two quantities for fast and slow wind and for magnetic field and
velocity will turn out to be useful to better interpret the radial evolution of intermittency as observed in the
solar wind MHD turbulence. Our analysis will be based on the following definitions: 1) a given time series is
defined intermittent if the factor $\mathcal{F}$ monotonically increases moving from larger to smaller scales, 2)
the same time series is defined more intermittent than another one if $\mathcal{F}$ begins to increase at larger
scales since, following Castaing et al. (1990), this implies a larger inertial range and, consequently, a larger
number of steps along the cascade with intermittency increasing at each step, 3)if $\mathcal{F}$ starts to
increase at the same scale for two different time series, we will consider more intermittent the one for which
$\mathcal{F}$ grows more rapidly. Moreover, we like to remind that a Gaussian statistics would show values of
$\mathcal{F}$ close to 3 for all scales, indicating the self--similar character of our fluctuations. However, if
$\mathcal{F}$ fluctuates around a value somewhat different from 3 our fluctuations are still self--similar
although not Gaussian. Anyhow, in both cases these fluctuations are not considered intermittent.

Thus, our definition of intermittency will be limited to a
qualitative definition rather than quantitative since the aim of
the present work is only to compare the radial evolution of
intermittency for different solar wind parameters and within
different solar wind conditions.

\section{3. Magnetic field and velocity intermittency vs heliocentric distance}

Values of $\mathcal{F}$ for both scalar and vector differences for magnetic field as a function of temporal scale
$\tau$ expressed in seconds are shown in  Figure~2. The factor $\mathcal{F}$ has been computed for slow (left
column) and fast (right column) wind and for three distinct radial distances as indicated by the different symbols
used in the plots. In addition, errors relative to each value of $\mathcal{F}$ are also shown. It is readily seen
that magnetic field fluctuations in both slow and fast wind are intermittent since $\mathcal{F}$ increases at
small scales. Values of $\mathcal{F}$ for compressive fluctuations within slow wind (panel A) start to increase
well beyond $10^4$ sec reaching values larger than 20 at the smallest scale. Moreover, there is no evidence for
any radial dependence since all the curves overlap to each other within the error bars. On the contrary,
compressive fluctuations for fast wind (panel B) show a clear radial dependence. As a matter of fact, the three
curves intersect each other at large scales down to $\sim 4\cdot10^3$ sec but clearly separate at smaller scales
indicating that intermittency increases with the radial distance from the sun. Moreover, since within slow wind
$\mathcal{F}$ starts to grow at larger scales and reaches higher values at small scales, we can say that magnetic
compressive fluctuations in slow wind, at least at 0.3 and 0.7 AU, are more intermittent than those within fast
wind. Slow wind directional fluctuations (panel C) are also rather intermittent since $\mathcal{F}$ starts to
increase around $2\cdot10^4$ sec, at frequencies slightly higher than for slow wind compressive fluctuations.
However, these fluctuations are less intermittent than compressive fluctuations in the same type of wind since
$\mathcal{F}$ increases more slowly at small scales. Moreover, there is no radial dependence. Panel D shows the
behavior of $\mathcal{F}$ for directional fluctuations in fast wind. Also in this case as for panel B, there is a
clear radial dependence of intermittency on the radial distance. The flatness factor $\mathcal{F}$ remains
approximately constant and rather similar for the three distances at large scales down to $\sim 2\cdot10^3$ sec
and then increases more rapidly for larger heliocentric distances. Thus, our sample at 0.9 AU is more intermittent
than that at 0.7 AU, which, in turn, is more intermittent than that at 0.3 AU. Considering that the scales at
which $\mathcal{F}$ starts to increase is only around $10^3$ sec and that the values reached at small scales are
lower, these fluctuations are less intermittent than the corresponding ones within slow wind and the compressive
ones within the same fast wind. Moreover, the fact that in panel D $\mathcal{F}$ starts to increase at much
smaller scales than in slow wind (Panel C), is strongly indicative that the inertial range in this case is much
less extended as we already know from the existing literature (see review by Tu and Marsch, 1995).

Results relative to velocity fluctuations are shown in Figure~3 in the same format adopted in the previous Figure.
Values of $\mathcal{F}$ for slow speed compressive fluctuations start to increase around $10^4$ sec (panel A).
However, the three curves intersect each other various times along the whole range of scales showing that there is
no clear radial dependence although, the smallest scale would suggest some radial evolution which, in addition,
would be opposite to what is observed in fast wind. However, the large associated error bars do not allow us to
draw any realistic conclusion. Taking into account that the the scale at which $\mathcal{F}$ starts to increase is
of the same order of that relative to magnetic field compressive fluctuations in slow wind but $\mathcal{F}$
reaches much lower values at small scales, we conclude that velocity compressive fluctuations are less
intermittent than magnetic compressive fluctuations in slow wind. Panel B, relative to velocity compressive
fluctuations in fast wind, shows a clear radial dependence of $\mathcal{F}$. The three curves start to increase
around $10^3$ sec and separate at smaller scales. Then, intermittency increases from 0.3 to 0.9 AU since
$\mathcal{F}$ increases more rapidly for larger heliocentric distances. This result is similar to what we observed
for magnetic compressive fluctuations in fast wind although the overall intermittency in this case is much reduced
taking into account that $\mathcal{F}$ starts to increase at much smaller scales and reaches smaller values. Panel
C shows results relative to velocity directional fluctuations in slow wind. These curves, although less stable
than the corresponding curves relative to magnetic field (Figure 2C),  show a very similar behavior and no hint
for a possible radial dependence. Also panel D, where we report values of $\mathcal{F}$ for velocity directional
fluctuations in fast wind, shows results very similar to those shown for magnetic field in Figure 2D to the extent
that these two sets of curves overlap to each other, within the error. This last result, as it will be discussed
later on in this paper, clearly derives from the strong contribution due to Alfv\'{e}nic fluctuations populating
the fast corotating streams that we selected (Bruno et al., 1985). Differently from magnetic field fluctuations,
velocity directional fluctuations seem to be only slightly less intermittent than compressive fluctuations.

\section{4. Intermittency in the mean field reference system}

Although, other authors (Marsch and Liu, 1993, Marsch and Tu, 1994) already addressed the study of the radial
evolution of intermittency for magnetic field and velocity components, we like to provide a complete picture of
this radial dependence adding a study performed in the mean field coordinate system (MF, hereafter) which, for
magnetic field, is more appropriate than the usual RTN or SE coordinate systems. As a matter of fact, the large
scale interplanetary magnetic field configuration breaks the spatial symmetry and introduces a preferential
direction along the mean field. As a consequence, a natural reference system is the one for which one of the
components, that we call $B_{//}$, is along the mean field $\vec{B}$ outwardly oriented, and the other two are
perpendicular to this direction. In our case we chose one of the two perpendicular components $B_{\bot 2}$ to be
perpendicular to the plane identified by $\vec{B}$ and the mean solar wind velocity $\vec{V}$, so that
$\hat{B}_{\bot 2}=\hat{\vec{B}}\times\hat{\vec{V}}$ and, the remaining direction $\hat{B_{\bot 1}}$ descends from
the vector product $\hat{B}_{\bot 2}\times\hat{\vec{B}}$,
where the symbol $\hat{}$ indicates a unitary vector. In the top panel of Figure~4, we show values of
$\mathcal{F}$ for the three magnetic components in SE reference system (i.e. $B_X$, $B_Y$ and $B_Z$) at the three
different heliocentric distances previously chosen and, in the bottom panel, we show the components in the MF
reference system (i.e. $B_{//}$, $B_{\bot 1}$ and $B_{\bot 2}$)for the same heliocentric distances. In the top
panel $\mathcal{F}$ increases for all the components as the radial distance increases. While at 0.9 AU the three
components show the same behavior, at 0.3 and 0.7 AU the curve relative to $B_X$ runs above the other two curves.
Since the distance between these curves slightly increases at small scales we might conclude that the radial
component is slightly more intermittent than the other two components. Unfortunately, this conclusion is not
corroborated by the size of the errors associated to each point, which are quite large. The bottom panel shows
results relative to the MF reference system. At 0.3
 and 0.7 AU there is not much difference with the situation
discussed in the previous panel because the orientation of the two reference systems is not very different either,
given that the magnetic field is almost radially oriented (see Table 1). On the contrary, moving to 0.9 AU and
comparing these results with those obtained in the other reference system, we clearly observe a decrease of
$\mathcal{F}$ for the two perpendicular components and an increase for the parallel component. Since the three
curves lie on the same level at large scales and end up with remarkable different values at small scales, we
conclude that the component parallel to the local magnetic field is more intermittent than the perpendicular
components. Moreover, the two perpendicular components are less intermittent in the MF reference system than in
SE. This means that in the MF reference system we enhance on one hand the stochastic character of the fluctuations
perpendicular to the local field direction and, on the other hand, the coherent character of the fluctuations
along the local field direction.

In Figure~5 we show, for the slow wind, the same elements discussed in the previous Figure. The much steeper
behavior of these curves suggests that, generally, slow wind is more intermittent than fast wind. Moreover,
especially at 0.3 AU, there is a tendency for both $B_X$, in the top panel, and $B_{//}$, in the bottom panel, to
be steeper than the other components at small scales,  suggesting a higher intermittency but, this tendency is not
confirmed at larger heliocentric distances. As discussed in the following, the reason for this appreciable
different behavior of $B_X$ and $B_{//}$ might be due to the fact that so close to the sun the contribution due to
Alfv\'{e}nic fluctuations, mainly acting on the perpendicular components, is not negligible even within slow wind
(Bruno et al., 1991). As a matter of fact, the stochastic nature of the fluctuations due to Alfv\'en waves tends
to make more Gaussian the PDFs of magnetic and velocity fluctuations perpendicular to the mean field direction. In
conclusion, an overall view reveals that the behavior of $\mathcal{F}$ within slow wind is not very sensitive to
this change of reference system.

For sake of completeness we have rotated into the MF reference system also velocity fluctuations although this
reference system is not the most appropriate for this parameter given that the wind expands radially. The two
panels of  Figure~6 show, for the three heliocentric distances, the behavior of $\mathcal{F}$ for fast wind
velocity fluctuations in the SE reference system and in the MF reference system, respectively. We like to remark
that in the SE reference system $V_X$ resembles very closely the behavior of the fast wind speed shown in Figure~3
since the average wind velocity vector is always close to the radial direction. Moreover, the two perpendicular
components, $V_Y$ and $V_Z$ in SE and $V_{\bot 1}$ and $V_{\bot 2}$ in MF, at 0.3 and at 0.7 AU closely recall the
behavior of the corresponding magnetic components within fast wind. However, the presence of a large plateau in
the central part of $V_X$ and $V_{//}$ makes it more difficult to estimate the degree of intermittency of these
components with respect to the perpendicular ones. At 0.9 AU, due to a weaker stationarity in the data, the
situation looks even more complex and does not allow to estimate which component is the most intermittent one.
$\mathcal{F}$ remarkably increases with distance at small scales for all the components in both reference systems.
However, as expected, the rotation into the MF reference system does not have a large influence at 0.3 AU but it
causes a general increase of $\mathcal{F}$ at  0.9 AU. The enhancement is such that the two perpendicular
components have the same behavior, and differences with the parallel component become appreciably smaller. This is
due to the fact that in this reference system the fluctuations of the components are not longer independent from
each other as it would be in SE reference system.

Finally, in Figure~7, we show results relative to the slow wind in the same format of the previous Figure. Here,
the very confused behavior of the curves and the large associated errors, especially at 0.7 and 0.9 AU, suggest a
rather weak stationarity of the data and make it difficult to compare the behavior of different components. A
general comment that we can easily make is that these curves are much steeper and start to increase at much larger
scales than in fast wind. As a consequence, velocity components in slow wind are generally more intermittent than
in fast wind. In addition, the rotation from SE to MF does not influence much our results, as expected. However,
it is such that the behavior of the three velocity components at 0.3 AU looks more similar to that of the
corresponding magnetic components (Figure~5, lower panel). This suggests that Alfv\'en waves, although less
relevant than in fast wind, might play a role even in this sample of slow wind.

\section{5. Summary and discussion}

We studied the radial dependence of solar wind intermittency looking at magnetic field and velocity fluctuations
between 0.3 and 1 AU. In particular, we analyzed  compressive and directional fluctuations for both fast and slow
wind. Our analysis exploits the property that probability distribution functions of a fluctuating field affected
by intermittency become more and more peaked at smaller and smaller scales. Since the peakedness of a distribution
is measured by its flatness factor we studied the behavior of this parameter at different scales to estimate the
degree of intermittency of our time series. Our general results can be summarized in the following points:

1) magnetic field fluctuations are more intermittent than velocity
fluctuations;

2) compressive fluctuations are more intermittent than
directional fluctuations;

3) slow wind intermittency does not show radial dependence;

4) fast wind intermittency, for both magnetic field and velocity,
clearly increases with distance.

5) magnetic and velocity fluctuations have a rather Gaussian behavior at large scales, as expected, regardless of
type of wind or heliocentric distance.

Point 4 is particularly interesting because we found that both
compressive and directional fluctuations become more intermittent
with distance. As a matter of fact, if we think of relations
\ref{compressive} and \ref{directional} we easily realize that
while intermittency of directional fluctuations can be fully
uncompressive, it is not possible to avoid that intermittency of
compressive fluctuations contaminates directional fluctuations.
In the latter case, the limiting condition would be the same
intermittency level for both kind of fluctuations. Thus,
intermittency of directional fluctuations contains also
contributions due to compressive fluctuations. This distinction
plays an important role in discussing our results since the
intermittency character of directional fluctuations reflects the
contribution of both compressive phenomena and uncompressive
fluctuations like Alfv\'{e}n waves.

Now, there are at least two questions that we should address: 1)
why directional fluctuations are always less intermittent than
compressive fluctuations? and, 2) why only fast wind shows radial
evolution? We can explain our observations simply assuming that
the two major ingredients of interplanetary MHD fluctuations are
compressive fluctuations due to a sort of underlying, coherent
structure convected by the wind and stochastic Alfv\'{e}nic
fluctuations propagating in the wind. The coherent nature of the
first ingredient would contribute to increase intermittency while
the stochastic character of the second one would contribute to
decrease it. If this is the case, coherent structures convected
by the wind would contribute to the intermittency of compressive
fluctuations and, at the same time, would also produce
intermittency in directional fluctuations. However, since
directional fluctuations are greatly influenced by Alfv\'{e}nic
stochastic fluctuations, their intermittency will be more or less
reduced depending on the amplitude of the Alfv\'en waves with
respect to the amplitude of compressive fluctuations. Thus,
compressive fluctuations would always be more intermittent than
directional fluctuations.

Before addressing the second question, we like to recall that several papers have already shown (see review by Tu
and Marsch, 1995) that slow wind Alfv\'enicity does not evolve with increasing the radial distance from the sun.
As a matter of fact, power spectra exhibit a spectral index close to that of Kolmogorov and a rather good
equipartition between \textit{inward} and \textit{outward} modes (Tu et al., 1989). Thus, once the inertial range
is established, the Alfv\'enicity of the fluctuations freezes to a state that, successively, is convected by the
wind into the interplanetary space without major changes (Bavassano et al., 2001). On the other hand, within fast
wind, turbulence is dominated by outward propagating Alfv\'en waves, which strongly evolve in the inner
heliosphere becoming weaker and weaker during the wind expansion, to the extent that at 1 AU, on the ecliptic
plane, their amplitude is much reduced and of the order of that of inward propagating Alfv\'en waves. At that
point, the resulting Alfv\'enicity resembles the one already observed in the slow wind close to the sun (Tu and
Marsch, 1990). Keeping this in mind, taking into account that convected structures experience a much slower radial
evolution because they do not interact with each other non--linearly as Alfv\'{e}n waves do, considering that
Alfv\'{e}n waves are mainly found in fast rather than in slow wind, it comes natural to expect that intermittency
would radially evolve within fast rather than slow wind. Obviously, this would explain why directional
fluctuations become more intermittent only within fast wind but would not explain why also compressive
fluctuations become more intermittent within fast wind. In reality, if we consider that compressive events cause
intermittency (Veltri and Mangeney, 1991, Bruno et al., 2001), we might ascribe this different behavior to the
fact that fast wind becomes more and more compressive with radial distance while the compressive level of slow
wind remains approximately the same, as shown by Marsch and Tu, (1990).

Our analysis performed on the components can also help to
understand, although partially, the topology of these convected
structures. In SE reference system, fluctuations along the radial
component are more intermittent than those perpendicular to it as
already found by Marsch and Liu (1993), although this feature,
especially for the magnetic field, tends to vanish around 1 AU.
The reason is that perpendicular components are more influenced
by Alfv\'{e}nic fluctuations and as a consequence their
fluctuations are more stochastic and less intermittent. This
effect largely reduces during the radial excursion mainly because
the SE reference system is not the most appropriate one for
studying magnetic field fluctuations. As a matter of fact, the
presence of the large scale spiral magnetic field breaks the
spatial symmetry introducing a preferential direction parallel to
the mean field. Consequently, we showed, that if we rotate our
magnetic data into the mean field reference system, especially at
0.9 AU, the intermittency of the perpendicular components
decreases and that of the parallel component increases. Moreover,
the two perpendicular components show a remarkable similar
behavior as expected if they experience Alfv\'{e}nic
fluctuations. On the other hand, results obtained on velocity
fluctuations suggest that a reference system with an axis
parallel to the radial direction looks more appropriate to
perform a similar study showing that the radial component seems
to be the most intermittent component.

One further observation is that generally most of the curves relative to velocity fluctuations came out to be less
stable than those relative to magnetic field fluctuations and affected by larger errors due to a weaker
stationarity with respect to magnetic field data.

Finally, our results cannot establish whether magnetic and velocity structures causing intermittency are convected
directly from the source regions of the solar wind or they are locally generated by stream--stream dynamic
interaction or, as an alternative view would suggest (Primavera et al., 2002), they are locally created by
parametric decay instability of large amplitude Alfv\`en waves. Probably all these origins coexist at the same
time and confirm that in any case the the radial dependence of the intermittency of interplanetary fluctuations is
strongly related to the turbulent radial evolution of their spectrum.

\vskip 1.5truecm

\acknowledgements We thank F. Mariani and N. F. Ness, PI's of the magnetic experiment and, H. Rosenbauer and R.
Schwenn, PI's of the plasma experiment onboard Helios 1 and 2, for allowing us to use their data. We also thank
both Referees for their valuable comments and suggestions.

\end{article}

\clearpage

\begin{table}
\centering
\begin{tabular}{|c|c|c|c|c|c|}
\hline time interval & radial distance [AU] & $<\mathbf{V}>
[km/s]$&  $\hat{\alpha}_{B-V}[^{\circ}]$ &
$\hat{\beta}_{B-R}[^{\circ}]$
&  $\hat{\gamma}_{V-R}[^{\circ}]$  \\
\hline
46:00--48:00 & 0.90 & 433 & 29.4 & 29.6 & 2.9 \\
49:12--51:12 & 0.88 & 643 & 31.4 & 29.6 & 2.4 \\
72:00--74:00 & 0.69 & 412 & 15.2 & 16.3 & 2.9 \\
75:12--77:12 & 0.65 & 630 & 19.4 & 18.1 & 1.5 \\
99:12--101:12 & 0.34 & 405 & 22.9 & 20.6 & 2.3 \\
105:12--107:12 & 0.29 & 729 & 8.2 & 7.5 & 1.8 \\
 \hline
\end{tabular}
\caption{From left to right: time interval in dd:hh, heliocentric distance in AU, average wind velocity in km/s,
angle between field and velocity vectors, angle between field vector and radial distance and, finally, angle
between velocity vector and radial distance.}
\end{table}

\clearpage

\section{Figures}
\begin{figure}
\epsfxsize=15.0cm \hspace{0.5cm} \epsfbox{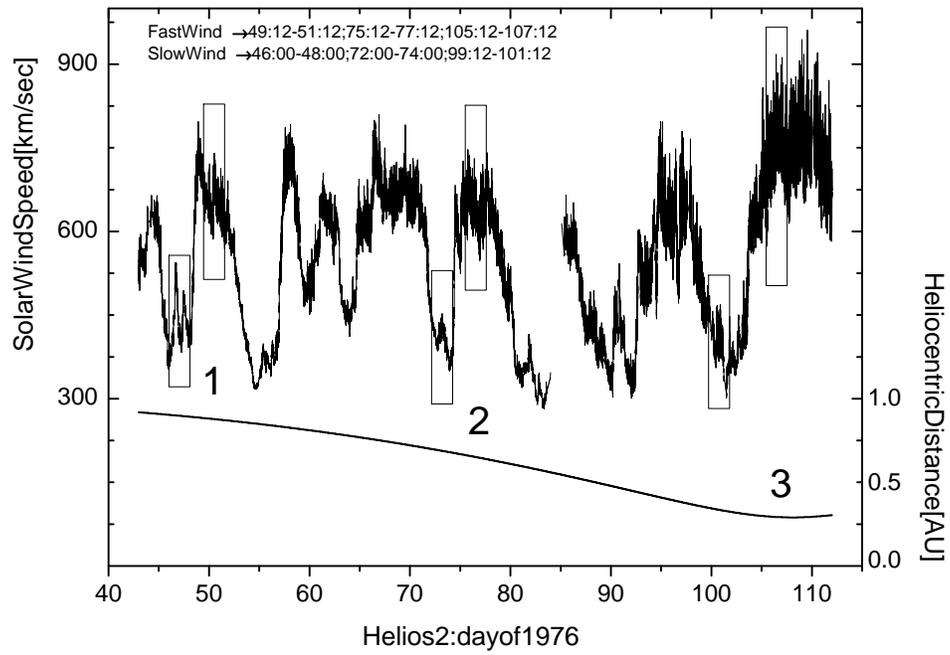}
\caption{Solar wind speed profile during Helios
2 primary mission. Rectangles overlaying the plot indicate the time intervals selected for the analysis (see also
Table 1 for more details on the intervals).}
\end{figure}
\vskip 12pt

\begin{figure}
\epsfxsize=15.0cm \hspace{0.5cm} \epsfbox{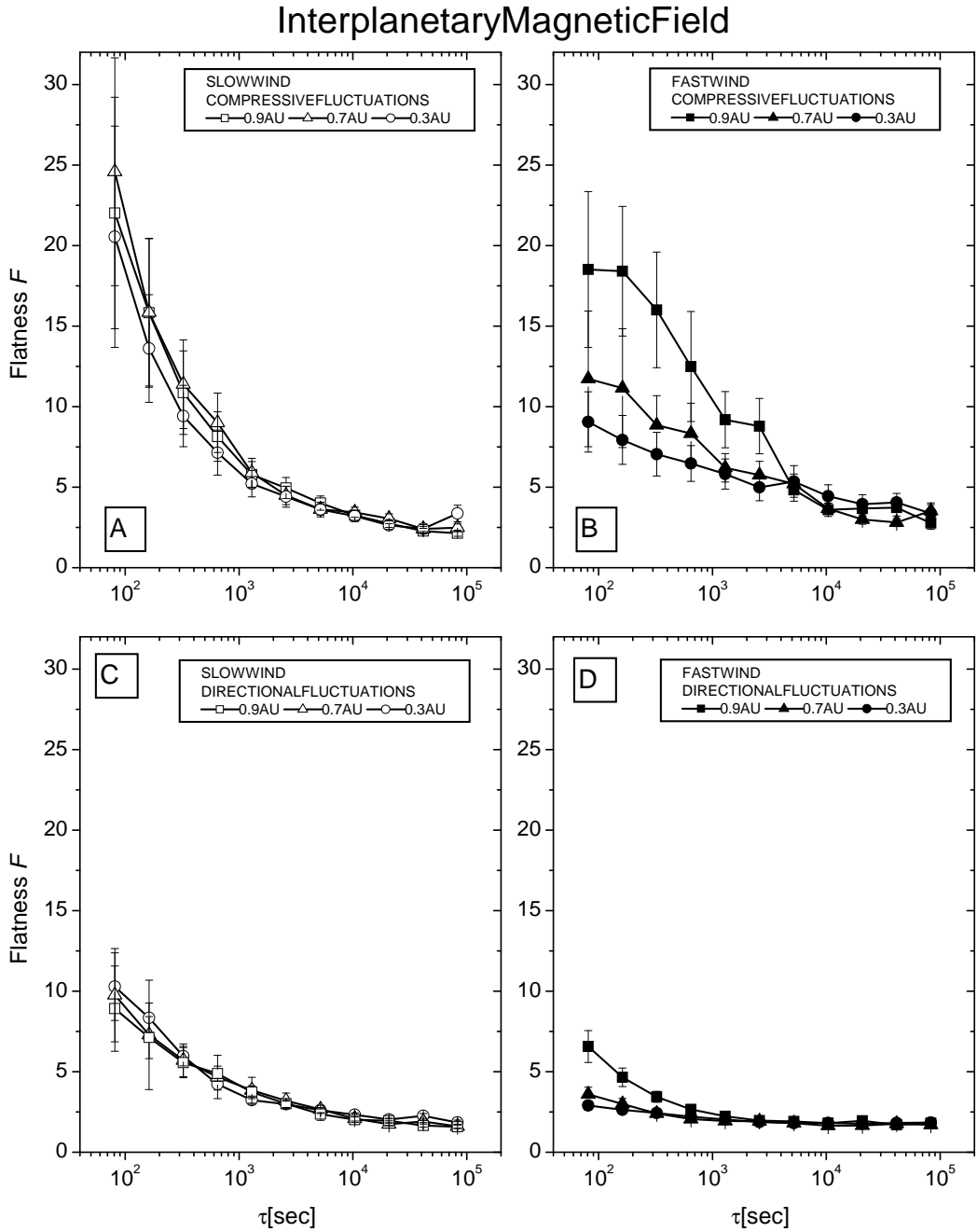}
\caption{Flatness $\mathcal{F}$ versus time
scale $\tau$ relative to magnetic field fluctuations. The left column (panels A and C) refers to slow wind and the
right column (panels B and D) refers to fast wind. The upper panels refer to compressive fluctuations, the lower
panels refer to directional fluctuations. Vertical bars represent errors associated to each value of
$\mathcal{F}$. The three different symbols in each panel refer to different heliocentric distances as reported in
the legend.}
\end{figure}
\vskip 12pt

\begin{figure}
\epsfxsize=15.0cm \hspace{0.5cm} \epsfbox{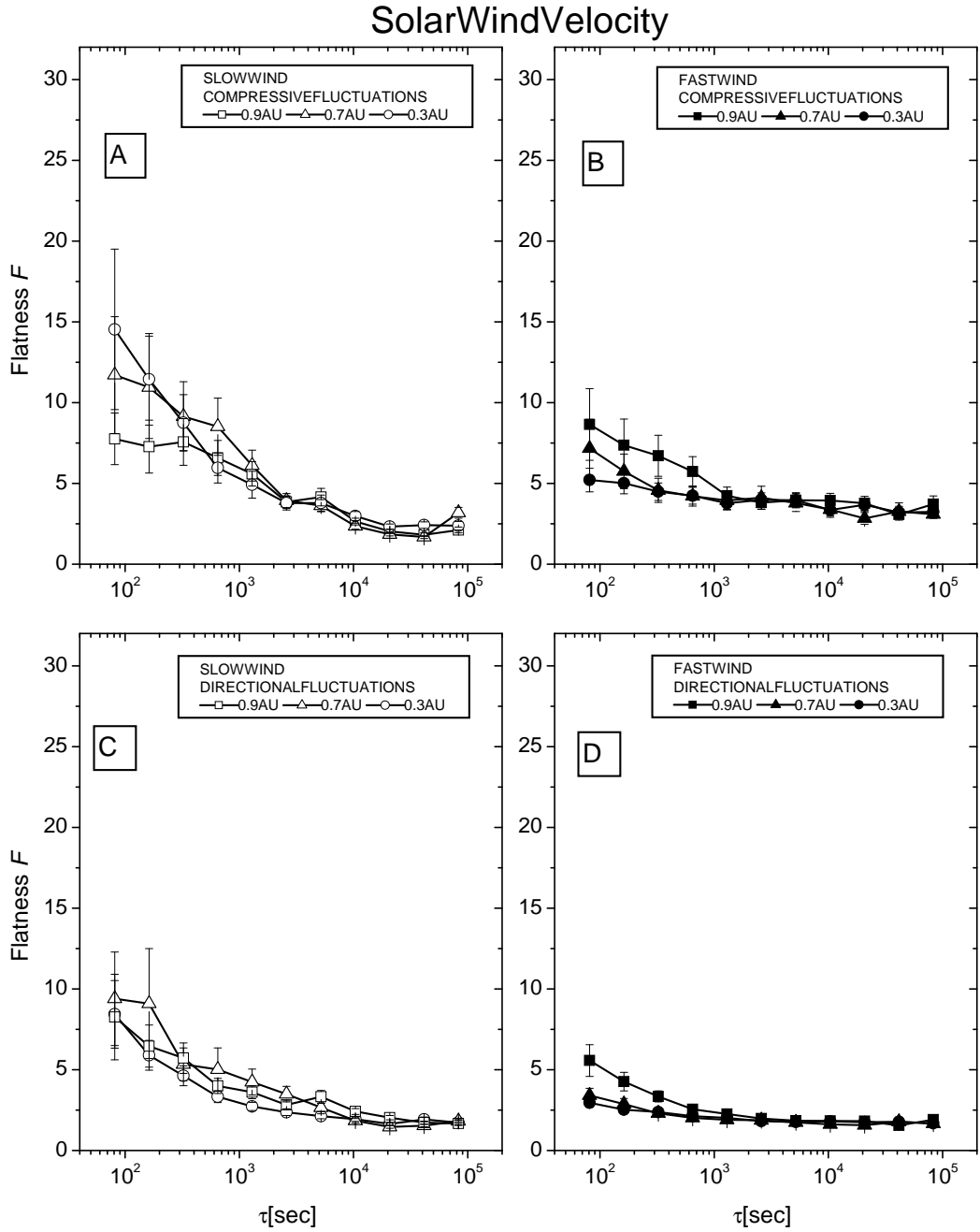}
\caption{Flatness $\mathcal{F}$ versus time
scale $\tau$ relative to wind velocity fluctuations. In the same format of Figure~2 panels A and C refer to slow
wind and panels B and D refer to fast wind. The upper panels refer to compressive fluctuations and the lower
panels refer to directional fluctuations. Vertical bars represent errors associated to each value of
$\mathcal{F}$. }
\end{figure}
\vskip 12pt

\begin{figure}
\epsfxsize=15.0cm \hspace{0.5cm} \epsfbox{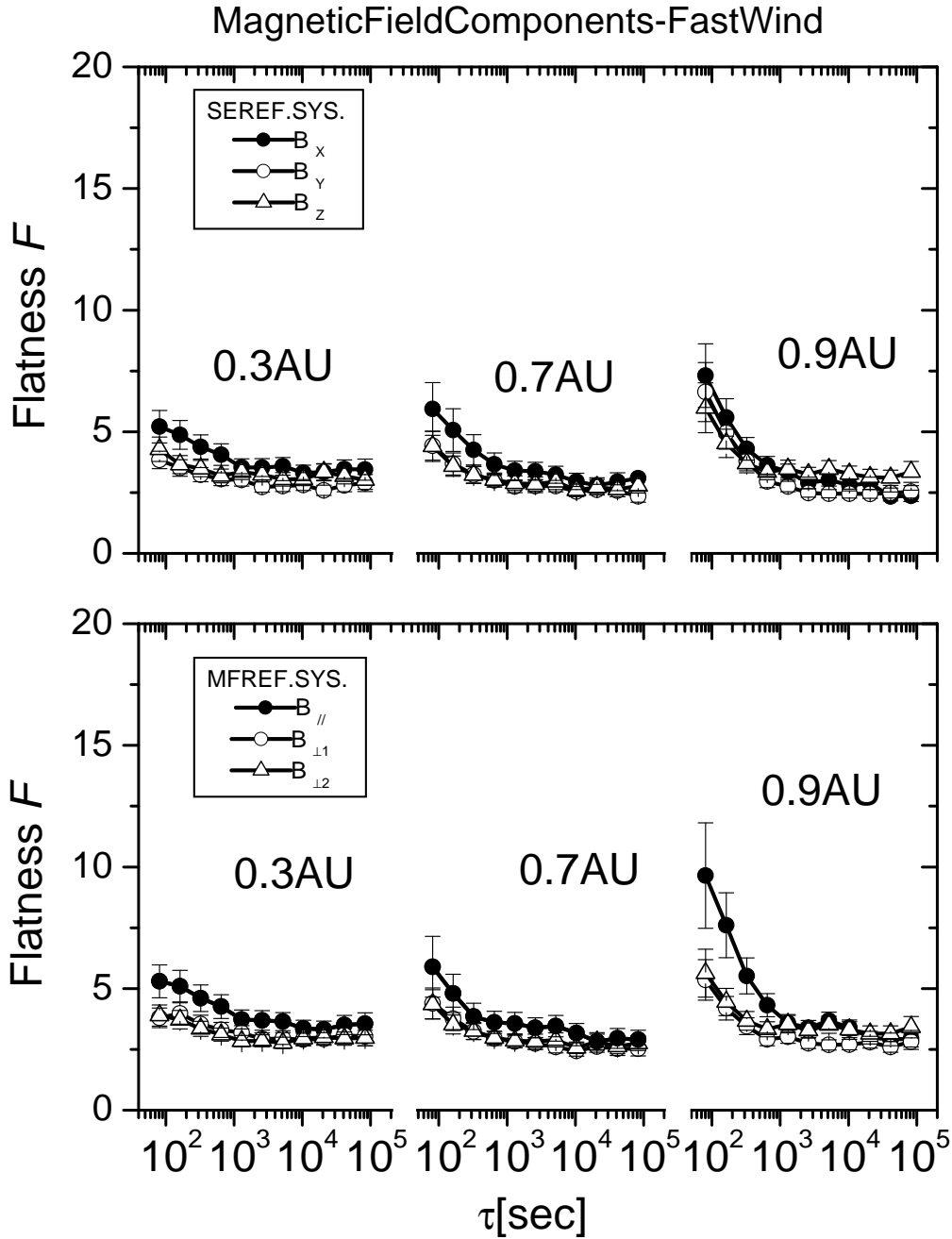}
\caption{Flatness $\mathcal{F}$ versus time
scale $\tau$ relative to fluctuations of the components of the interplanetary magnetic field observed in fast
wind. The scale of the horizontal axis is divided in three parts all covering the same range of scales. \\ Upper
panel) there are three sets of curves at three different heliocentric distance. Within each set, different
components are indicated by different symbols as reported in the legend. Components in this panel are taken in the
Solar Ecliptic (SE) reference system where, the $\mathcal{X}$ axis is oriented towards the sun, the $\mathcal{Y}$
axis lies on the ecliptic and it is oriented opposite to the s/c direction of motion and, the $\mathcal{Z}$ axis
completes the right--handed reference system. \\ Lower panel) parallel and perpendicular components in the Mean
Field (MF) reference system. $B_{//}$, is along the main field $\vec{B}$ outwardly oriented, $B_{\bot 2}$ is
perpendicular to the plane identified by $\vec{B}$ and the mean solar wind velocity $\vec{V}$, so that
$\hat{B}_{\bot 2}=\hat{\vec{B}}\times\hat{\vec{V}}$ and, the remaining direction $\hat{B_{\bot 1}}$ descends from
the vector product $\hat{B}_{\bot 2}\times\hat{\vec{B}}$. Different components are indicated by different symbols
as reported in the legend.}
\end{figure}
\vskip 12pt

\begin{figure}
\epsfxsize=15.0cm \hspace{0.5cm} \epsfbox{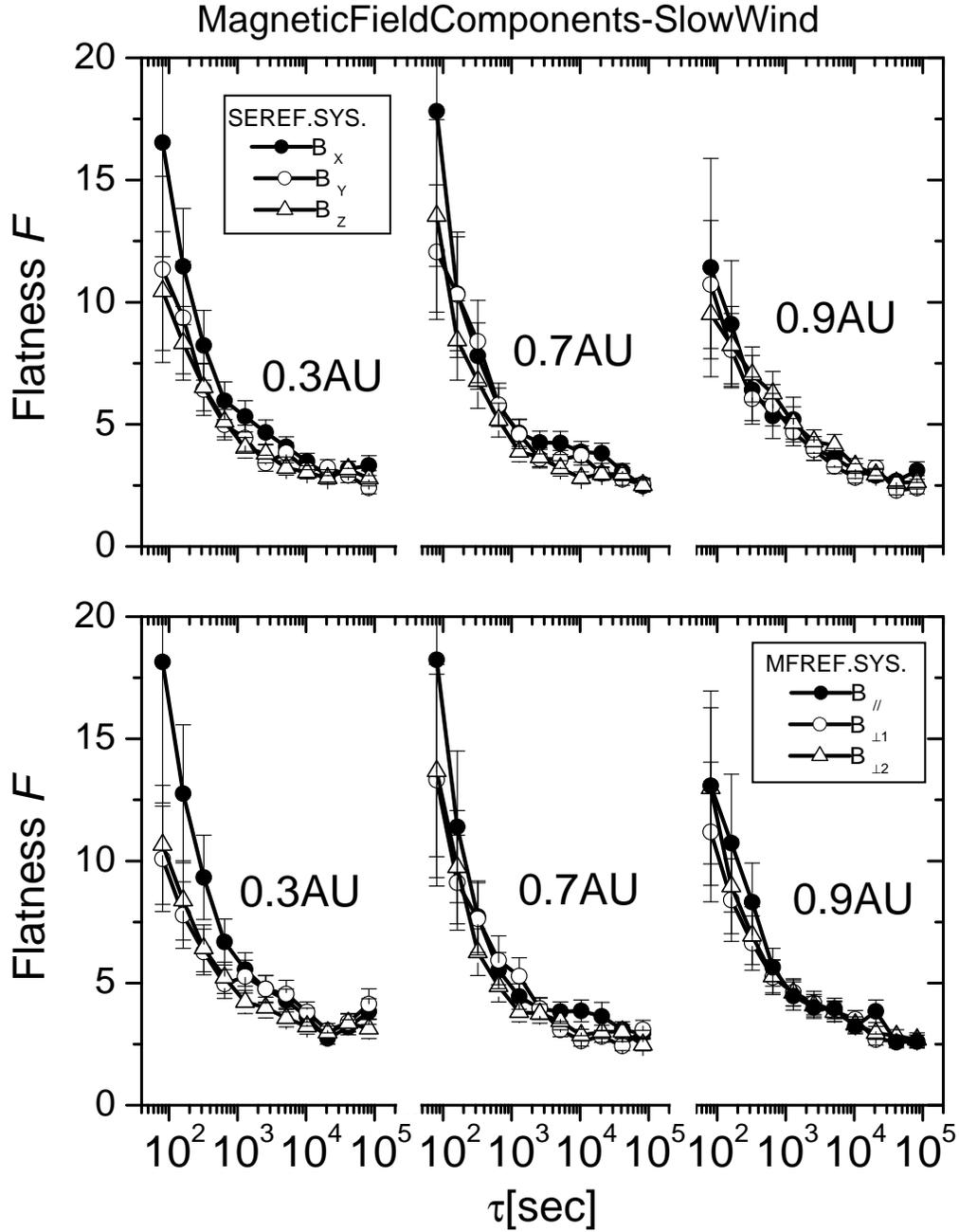}
\caption{Flatness $\mathcal{F}$ versus time
scale $\tau$ relative to fluctuations of the components of the interplanetary magnetic field observed in slow
wind, in the same format of Figure~4.}
\end{figure}
\vskip 12pt

\begin{figure}
\epsfxsize=15.0cm \hspace{0.5cm} \epsfbox{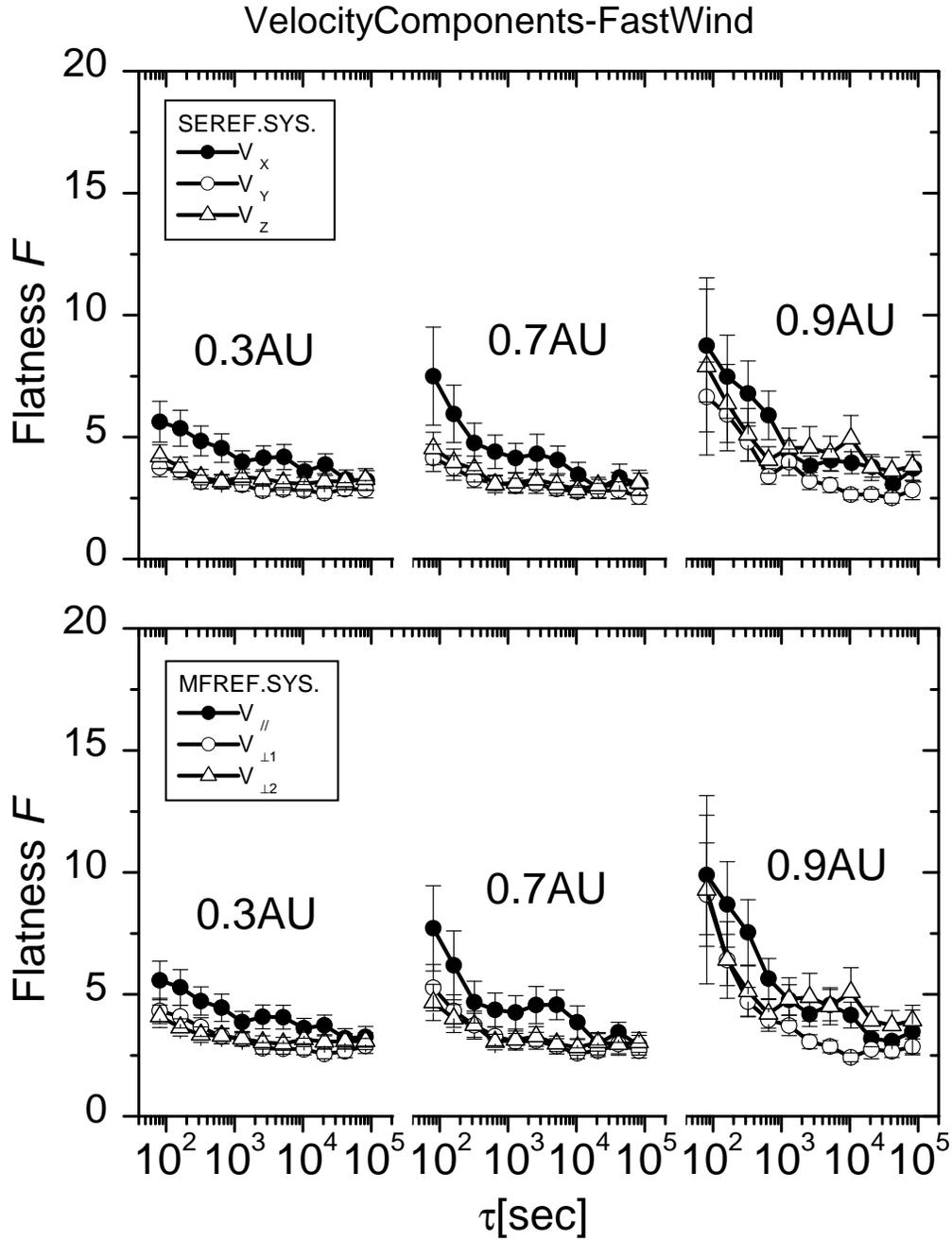}
\caption{Flatness $\mathcal{F}$ versus time
scale $\tau$ relative to fluctuations of velocity components observed within fast wind. Results in the upper panel
refer to components observed in the SE reference system while in the lower panel refer to solar wind components
rotated into the mean field (MF) reference system. Format and symbols are the same used for Figures~4 and 5.}
\end{figure}
\vskip 12pt

\begin{figure}
\epsfxsize=15.0cm \hspace{0.5cm} \epsfbox{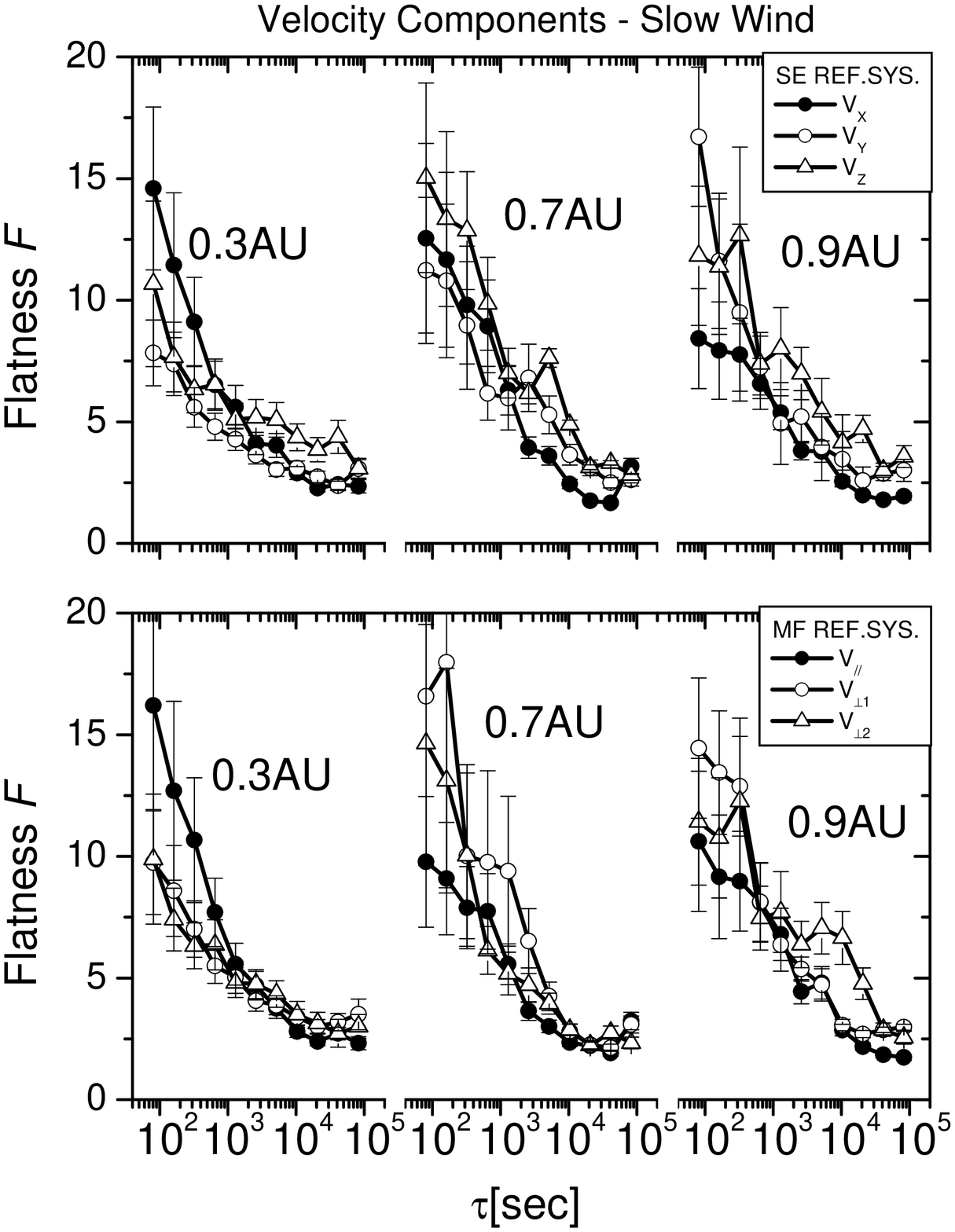}
\caption{Flatness $\mathcal{F}$ versus time
scale $\tau$ relative to fluctuations of velocity components observed in slow wind, in the same format of
Figure~6.}
\end{figure}


\begin{references}

\reference
Anselmet, F., Gagne, Y., Hopfinger, E. J., and Antonia, R.A., 1984,  High--order velocity structure
functionsin turbulent shear flows, J. Fluid Mech., 140, 63--89

\reference
Bavassano, B., M. Dobrowolny, F. Mariani, and N.F. Ness, 1982, Radial  evolution of power spectra of
interplanetary Alfv\'enic turbulence, J. Geophys. Res. 86, 3617--3622

\reference
Bavassano, B., M. Dobrowolny, G. Fanfoni, F. Mariani, and N.F. Ness, 1982, Statistical properties of MHD
fluctuations associated with high--speed streams from Helios--2 observations, Solar Phys., 78, 373--384


\reference
B. Bavassano, and R. Bruno, 1992, On the role of interplanetary  sources in the evolution of
low-frequency Alfv\'enic turbulence in the solar wind, J. Geophys. Res., 97, 19129--19137

\reference Bavassano, B., Pietropaolo, E., Bruno R., 2001, Radial evolution of outward and inward Alfv\'{e}nic
fluctuations in the solar wind: A comparison between equatorial and polar observations by Ulysses, J. Geophys.
Res., 106, 10659--10668

\reference Belcher, J. W., and L. Davis, Jr., 1971, Large--amplitude Alfv\'en waves in the interplanetary medium,
2,  J. Geophys. Res. 76, 3534--3563.

\reference
Benzi, R., Ciliberto, S., Tripiccione, R., Baudet, C, Massaioli, F., Succi, S., 1993, Extended
sel--similarity in turbulent flows, Phys. Rev. E 48, R29

\reference
Bruno, R., Bavassano, B. and Villante, U., 1985, Evidence for long period Alfv\'en waves in the inner
solar system, J. Geophys. Res. 90, 4373--4377.

\reference
Bruno, R. and Bavassano, B., 1991, Origin of low cross--helicity regions in the inner solar wind, J.
Geophys. Res. 96, 7841--7851.


\reference
Bruno, R., V. Carbone, P. Veltri, E. Pietropaolo and B. Bavassano, 2001, Identifying intermittent
events in the solar wind, Planetary Space Sci., 49, 1201--1210.

\reference Burlaga, L., 1991, Intermittent turbulence in the solar wind, J. Geophys. Res. 96, 5847--5851.

\reference
Carbone, V., 1993, Cascade model for intermittency in fully developed magnetohydrodynamic turbulence,
Phys. Rev. Lett. 71, 1546-1548.

\reference
Carbone,V., Veltri, P., Bruno, R., 1995, Experimental evidence for differences in the extended
self--silimarity scaling laws between fluid and magnetohydrodynamic turbulent flows, Phys. Rev. Lett. 75,
3110--3120.

\reference
Castaing, B., Gagne, Y., and Hopfinger, 1990, Velocity probability density functions of high Reynolds
number turbulence, Physica D 46, 177--200.

\reference
Coleman, P. J., 1968, Turbulence, viscosity and dissipation in the solar wind plasma, Astrophys. J.,
153, 371--388.

\reference
Denskat, K. U., and Neubauer, F. M., 1983, Observations of hydromagnetic turbulence in the solar wind,
Solar Wind V, edited by M. Neugebauer, NASA Conf. Publ., CP--2280, 81--91.

\reference
Frisch, U., 1995, Turbulence: the legacy of A. N. Kolmogorov, Cambridge University Press


\reference
Horbury, T. A., Balogh, A., Forsyth, R. J., Smith, E., 1996, Magnetic field signatures of unevolved
turbulence in solar polar flows, J. Geophys. Res.  101, 405--413.

\reference
L. Klein, R. Bruno, B. Bavassano and H. Rosenbauer, 1993, Anisotropy and minimum variance of
magnetohydrodynamic fluctuations in the inner heliosphere, J. Geophys. Res., 98, 17461--17466

\reference
Kolmogorov, A. N., 1941, The local structure of turbulence in incompressible viscous fluid for very
large Reynolds numbers, C. R. Akad. Sci. SSSR 30, 301.

\reference
Kolmogorov, A. N., 1962, A refinement  of previous hypotheses concerning the local structure of
turbulence in a viscous incompressible fluid at high Reynolds number, \textit{J. Fluid Mech.}, 13, 82--85

\reference
Kraichnan, R. H., 1965, Inertial--range spectrum of hydromagnetic turbulence, Phys. Fluids 8,
1385--1387.

\reference
Marsch, E. and Tu, C.-Y, 1990, On the radial evolution of MHD turbulence in the inner heliosphere, J.
Geophys. Res. 95, 8211--8229.

\reference
Marsch, E., and Liu, S., 1993, Structure functions and intermittency of velocity fluctuations in the
inner solar wind, Ann. Geophysicae 11, 227--238.

\reference
Marsch, E. and Tu, C.-Y, 1994, Non--Gaussian probability distributions of solar wind fluctuations, Ann.
Geophysicae 12, 1127--1138.

\reference
Matthaeus, W. H., Goldstein, M. L., 1982, Measurements of the rugged invariants of magnetohydrodynamic
turbulence in the solar wind, J. Geophys. Res., 87, 6011--6028

\reference
Matthaeus, W. H., Goldstein, M. L., Roberts, D. A., 1990, Evidence for the presence of
quasi--two--dimensional nearly incompressible fluctuations in the solar wind, J. Geophys. Res. 95, 20673--20683.

\reference
McCracken, K.G., and N.F. Ness, 1966, The collimation of cosmic rays by the interplanetary magnetic
field, J. Geophys. Res. 71, 3315--3332.

\reference
Meneveau, C., and Sreenivasan, K. R., 1987, Simple multifractal cascade model for fully developed
turbulence, Phys. Rev. Lett. 59, 1424--1427.

\reference
Obukhov, 1962, Some specific features of atmospheric turbulence, \textit{J. Fluid Mech.}, 13, 77--81

\reference
Padhye, N.~S., Smith, C.~W. and Matthaeus, W.~H.,2001, Distribution  of magnetic field components in
the solar wind plasma, J. Geophys. Res. 9106, 18635-18650.

\reference
Parisi, G. and, U. Frisch, 1985, On the singularity structure of fully developed turbulence, in
\textit{Turbulence and Predictability in Geophysical Fluid Dynamics, Proceed. Intern. School of Physics 'E.
Fermi', 1983, Varenna, Italy}, 84--87, eds. M. Ghil, R. Benzi and G. Parisi, North--Holland, Amsterdam

\reference
Primavera, L., F. Malara, and P. Veltri, 2002, Parametric instability  in the solar wind: numerical
study of the nonlinear evolution, paper presented at Solar Wind 10 Conference, 17--21 June 2002, Pisa, Italy

\reference
Roberts, D.A., M. Goldstein, L. Klein, and W. Matthaeus, 1987, Origin and evolution of fluctuations in
the solar wind: Helios observations and Helios-Voyager comparisons, J. Geophys. Res., 92, 12023--12035.

\reference
Roberts, D. A., M. L., Goldstein, W. H. Matthaeus and S. Gosh, 1991, MHD simulation of the radial
evolution and stream structure of solar wind turbulence, Phys. Rev. Lett., 67, 3741--3765.

\reference
Roberts, D. A.,1992, Observation and simulation of the radial evolution and stream structure  of solar
wind turbulence, in E. Marsch and R. Schwenn (eds) , Solar Wind Seven, COSPAR Colloquia Series Vol. 3, Pergamon
Press, Oxford, 533--538

\reference
Ruzmaikin, A., Feynman, J., Goldstein, B., and Balogh, A., 1995, Intermittent turbulence in solar wind
from the south polar hole, J. Geophys. Res. 100, 3395--3404.

\reference
Sorriso--Valvo, L. , Carbone, V., Veltri, P., Consolini, G., Bruno, R., 1999, Intermittency in the
solar wind turbulence through probability distribution functions of fluctuations, Geophys. Res. Lett. 26,
1801--1804.

\reference
Tu, C. -Y., Pu, Z.-Y, and Wei, F.-S, 1984, The power spectrum of interplanetary Alfv\'{e}nic
fluctuations: Derivation of its governing equation and its solution, J. Geophys. Res. 89, 9695--9702.

\reference
Tu, C. -Y., 1988, The damping of interplanetary Alfv\'enic fluctuations and the heating of the solar
wind, J. Geophys. Res. 93, 7--20.

\reference
Tu, C.-Y and Marsch, E., 1990, Evidence for a ``background'' spectrum of the solar wind turbulence in
the inner heliosphere, J. Geophys. Res. 95, 4337--4341.

\reference
Tu, C.-Y and Marsch, E., and Thime, K. M., 1989 Basic properties of solar wind MHD turbulence analysed
by means of Els\"{e}sser variables, J. Geophys. Res. 95, 11739--11759.

\reference
Tu, C.-Y and Marsch, E., 1993, A model of solar wind fluctuations with two components: Alfv\'en waves
and convective structures, J. Geophys. Res. 98, 1257--1276.

\reference
Tu, C.-Y and Marsch, E., 1995, MHD structures, waves and turbulence  in the solar wind: observations
and theories, Space Sci. Rev., 73, 1--210

\reference
Tu, C.-Y, Marsch, E., Rosenbauer, H., 1996, An extended structure  function model and its aplication to
the analysis of solar wind intermittency properties, Ann. Geophys. 14, 270--285.

\reference
Veltri, P., and Mangeney, A., 1999, in Solar Wind IX, edited by S. Habbal, AIP Conf. Publ., 543--546.

\end{references}
\end{document}